\documentclass[prx,amssymb,superscriptaddress, twocolumn]{revtex4-2}
\usepackage{amsmath}
\usepackage{amssymb}
\usepackage{subfigure}
\usepackage{graphicx}
\usepackage{ulem}
\usepackage{multirow}
\usepackage{enumitem}
\usepackage{hyperref}
\usepackage{appendix}
\usepackage{mathtools}
\usepackage{mathrsfs}
\usepackage{comment}

\newcommand{\mean}[1]{\langle #1\rangle}

\newcommand{\be}{\begin{equation}}
\newcommand{\ee}{\end{equation}}

\newcommand{\lf}{\left}
\newcommand{\rg}{\right}

\newcommand{\bea}{\begin{eqnarray}}
\newcommand{\eea}{\end{eqnarray}}

\newcommand{\nn}{\nonumber}

\newcommand{\opt}[1]{#1^*}
\DeclareMathOperator*{\argmax}{arg\,max}

\usepackage{xcolor}
\usepackage{color}
\newcommand{\change}[1]{{\color{black} {#1}}}

\begin{document}

\title{Optimal quantum \change{key distribution} networks:\\ capacitance
  \textit{versus} security }

\author{Lorenzo Cirigliano}
\affiliation{Dipartimento di Fisica Universit\`a ``Sapienza”, P.le
  A. Moro, 2, I-00185 Rome, Italy.}
\affiliation{Centro Ricerche Enrico Fermi, Piazza del Viminale, 1,
  I-00184 Rome, Italy}
\affiliation{Departamento de F\'\i sica da Universidade de Aveiro \& I3N, Campus Universit\'ario de Santiago, 3810-193 Aveiro, 
Portugal}
\author{Valentina Brosco}
\thanks{valentina.brosco@cnr.it}
\affiliation{Istituto dei Sistemi Complessi (ISC-CNR), Via dei Taurini
  19, I-00185 Rome, Italy}
\affiliation{Dipartimento di Fisica Universit\`a ``Sapienza”, P.le
  A. Moro, 2, I-00185 Rome, Italy.}
\affiliation{Centro Ricerche Enrico Fermi, Piazza del Viminale, 1,
  I-00184 Rome, Italy}

\author{Claudio Castellano}
\affiliation{Istituto dei Sistemi Complessi (ISC-CNR), Via dei Taurini
  19, I-00185 Rome, Italy}
\affiliation{Centro Ricerche Enrico Fermi, Piazza del Viminale, 1,
  I-00184 Rome, Italy}

\author{Claudio Conti}
\affiliation{Dipartimento di Fisica Universit\`a ``Sapienza”, P.le
  A. Moro, 2, I-00185 Rome, Italy.}
%\affiliation{Istituto dei Sistemi Complessi (ISC-CNR), Via dei Taurini
%  19, I-00185 Rome, Italy}  
%\affiliation{Centro Ricerche Enrico Fermi, Piazza del Viminale, 1,
%  I-00184 Rome, Italy}

\author{Laura Pilozzi}
\affiliation{Istituto dei Sistemi Complessi (ISC-CNR), Via dei Taurini
  19, I-00185 Rome, Italy}
\affiliation{Centro Ricerche Enrico Fermi, Piazza del Viminale, 1,
  I-00184 Rome, Italy}

\date{\today}
\begin{abstract}
The rate and security of quantum 
communications between users placed at arbitrary points of a quantum
communication network depend on the structure of the network, on its
extension and on the nature of the communication channels. In this
work we propose a strategy \change{for the optimization of trusted-relays based networks}
that intertwines classical network approaches and quantum information theory. Specifically,
by suitably defining a quantum \change{communication} efficiency functional,
we identify the optimal quantum communication connections through the
network by balancing security and the quantum
communication rate. The optimized network is then constructed as the
network of the maximal quantum \change{communication} efficiency connections and its performance
is evaluated by studying the scaling of average properties as functions
of the number of nodes and of the network spatial extension.
\end{abstract}

\maketitle

\section{Introduction}

Quantum communication networks \cite{azuma2020} enable the realization of tasks
beyond the reach of  classical communication systems. Examples are
unconditionally secure quantum key distribution
\cite{bennett1984,ekert1991} (QKD), quantum teleportation \cite{hermans2022},
clock-synchronization \cite{komar2014}, distributed quantum computing
\cite{wehner2018}, to mention just a few.
Characterizing and optimizing quantum communication networks have
a crucial relevance for the development of quantum cryptography
applications \cite{pirandola2020} and \change{hold} the potential to advance our
understanding of fundamental quantum phenomena \cite{nokkala2023complex}, such as entanglement
percolation \cite{acin2007} or the emergence of non-local quantum
correlations\cite{gisin2019,poderini2020,carvacho2022}.

The performance of quantum networks is determined by the nature of
the quantum communication channels and
protocols~\cite{pirandola2019,coutinho2022}
and by the overall network topology.
The optimization of quantum communication networks involves therefore
the closing of security loopholes and the mitigation of the effect of
losses through the development of quantum communication protocols,
such as for example the measurement device independent QKD
\cite{lo2012,pirandola2015,erkilic2023} and the twin-field QKD
\cite{lucamarini2018} protocols.
But it is also pursued by optimizing the allocation of quantum
resources for quantum sensing~\cite{krisnanda2023} and for distributed
quantum computing \cite{meter2007} or by engineering optimal routing
strategies \cite{meter2013,harney2022,hahn2019,schoute2016,pant2019},
taking into account the peculiar features of the network elements and
the network architecture.
 
In the ideal case the ultimate properties of network elements, such as
the quantum communication links, are dictated by the laws of quantum
mechanics, enforcing their security but also imposing intrinsic bounds \change{
\cite{pirandola2009,takeoka2014}} on the rate of quantum information transmission. 
 \change{Specifically, the fundamental limit of repeaterless quantum communication
found by Pirandola, Laurenza, Ottaviani
and Banchi \cite{pirandola2017}, known as  PLOB bound, prevents to achieve
simultaneously high rates and long distances in transferring quantum
states and distributing entanglement or secret quantum keys through a
quantum link. }

The global features of quantum communication networks are strongly
dependent on the spatial distribution of the users. Recent theoretical
works developed a random network approach to large-scale quantum
communication networks based on optical fibers~\cite{brito2020} or
satellite links~\cite{brito2021} and analyzed their connectivity,
nodes distance and the presence of small world features
\cite{brito2020,brito2021}.

In this work we employ the tools of classical network science to
devise a strategy of optimization of quantum communication networks.
PLOB bound can be indeed circumvented by means of intermediate
repeaters \cite{pirandola2019}, either of quantum~\cite{azuma2022} or
classical nature (trusted nodes~\cite{salvail2010,mehic2020}), that
help the communication between distant parties.  With few notable
exceptions~\cite{joshi2020}, most field tests of metropolitan-scale
quantum networks to date are based on point-to-point architecture and
they involve trusted nodes, see
Refs. \cite{elliott2005,peev2009,sasaki2011,wang2014,chen2021}.
Trusted nodes in general lower the security of the
network~\cite{scarani2009,salvail2010, solomons2022}.  Consequently,
in designing a QKD network the following question arises naturally:
given a set of QKD users, what is the optimal way to connect them to
fulfill a given rate/security target, assuming that all trusted nodes
have a certain probability $p$ of being leaky?  Here we address this
optimization problem by introducing a \textit{quantum \change{communication} efficiency
  functional}, that balances the quantum security and the quantum
communication rate for each pair of users in the network.
\change{
  Note that in a classical network leakage can occur not only at
  nodes where the signal is amplified, but also along the connections
  between nodes: adding amplifiers therefore improves the capacitance
  without necessarily reducing the security level. For this
  reason in classical networks there is no tradeoff between
  capacitance and security, which is instead inherent in QKD networks.}
We develop an algorithm that maximizes the quantum \change{communication} efficiency and
constructs the optimal network, that we refer to as maximal quantum \change{communication}
efficiency network.  We then investigate, for a random distribution of
users in the plane, the average properties of these optimal networks.
Their performance is evaluated by studying the scaling of average
properties as a function of the number of nodes and of the network
extension.  While the quantum communication rate is linked to
geometrical properties, such as the average distance between users,
the security depends on the topology.  The optimization algorithm
therefore goes beyond standard dynamic programming methods, such  as the
Dijkstra algorithm~\cite{dijkstra1959}, that \change{were} previously employed
in the context of quantum repeater network
optimization~\cite{meter2013}.
\section{Results}

\subsection{\change{Communication} efficiency of quantum networks}
\begin{figure*}[t]
    \includegraphics[width=\textwidth]{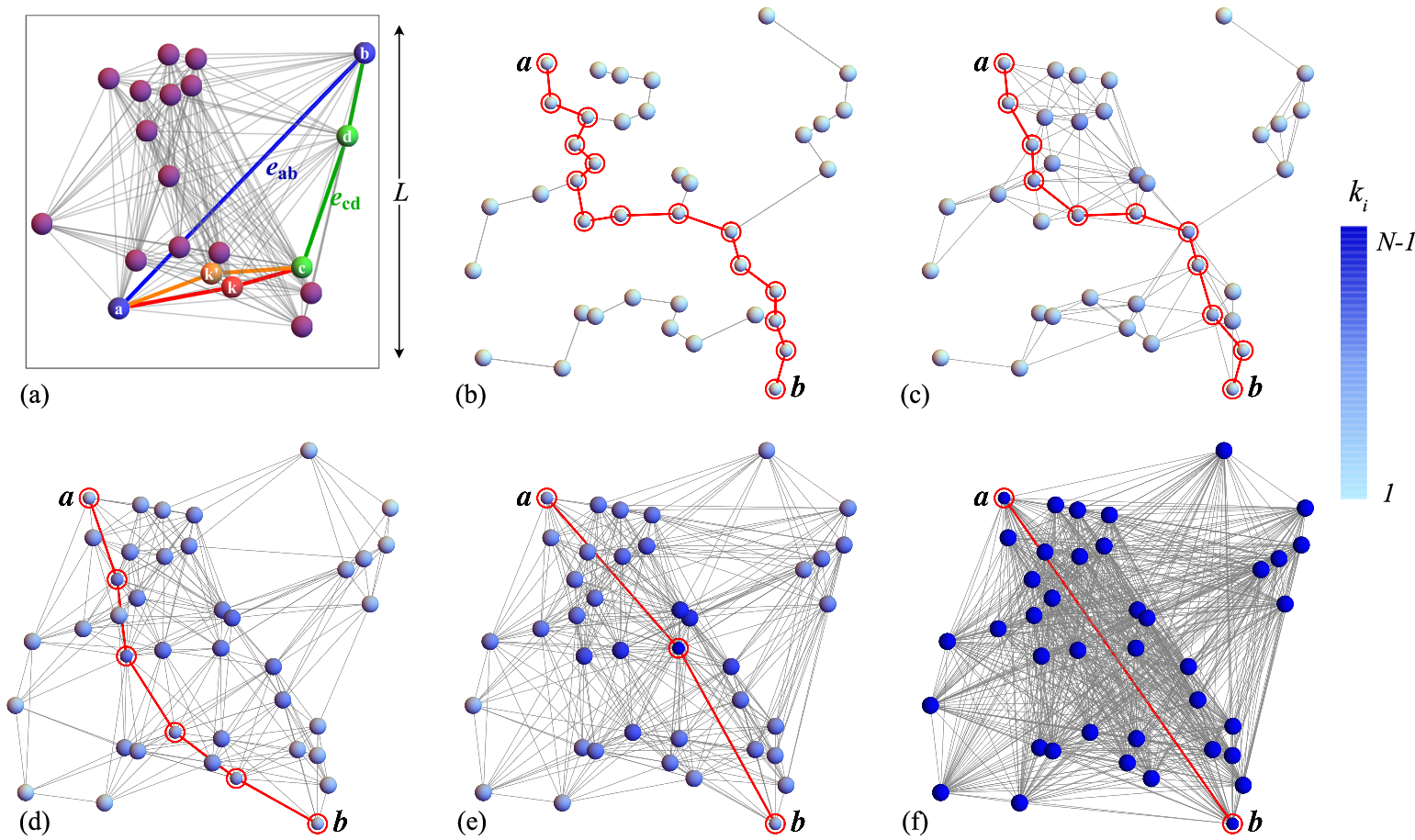}
  \caption{{\bf \change{The optimal path between two nodes.}}
    (a) Distribution of $N=20$ points in a square of size $L$ representing
    the users of a QKD network. The colored lines represent possible
    paths between users $a$ and $b$. (b) Maximum spanning tree for $N=40$ nodes.
    Here and in the following panels the red edges represent the optimal path
    between user $a$ and user $b$. Nodes are colored according to their degree.
    (c-f) Maximal Quantum \change{Communication} Efficiency networks ${\cal G}^*_\alpha$, obtained using
    our optimization algorithm,
    for $N = 40$, $L = \lambda_0$ and $\alpha = 0,\, 0.2,\, 0.3,\,1$.}
  \label{random-point}
\end{figure*}
In its simplest realization a QKD network consists of a set of $N$
users (nodes) that can send and receive quantum bits along a set of physical
links.
Here we assume that users are points located in a square of
side $L$, as shown in Fig.~\ref{random-point}.
As physical connections we consider lossy bosonic links. In this
case, following Ref.~\cite{pirandola2017}, the QKD rate of a link,
$e$, connecting two users located at the points ${\bf x}_a$ and ${\bf x}_b$ can be
quantified by its quantum capacitance $q(e)$
fulfilling the PLOB relation
\be
\label{chq}
      q(e)=-\log_2[(1 -e^{-d_{ab}/\lambda_0})].
\ee
where $d_{ab}=|{\bf x}_a-{\bf x}_b|$ is the Euclidean distance between the
users and $\lambda_0$ is a characteristic decay length. For optical
fibers the attenuation in the C telecom band is of the order of 0.2
db/Km yielding $\lambda_0 \sim 22$ Km.  Note that, since the link
capacitance provides an estimate of the number of qubits sent per use
of the channel, Eq.~\eqref{chq} sets to 15 Km the distance at which a
single qubit per use can be sent using a standard optical fiber
connection.
The quantum capacitance of a channel can be increased by
means of repeaters. In particular, connecting the users $a$ and $b$
through a path featuring $m$ trusted nodes,
the capacitance of the channel~\cite{pirandola2019} is given by
\begin{equation}\label{capacitance}
    q({\{a \to b\}}) = \min_{e \in \{a \to b\}}q(e).
\end{equation}
As an example let us consider users $a$ and $b$ shown in
Fig.~\ref{random-point}(a) and let us assume that all other users can act
as trusted nodes. We show two possible ways to connect $a$ and $b$: a direct
link or a path passing through three trusted nodes.
In the first case the quantum
capacitance is given by $q(e_{ab})$
while in the second case the capacitance is
$q(e_{cd}) > q(e_{ab})$.

Such an increased capacitance is however associated to a potential
vulnerability to attacks, since in most practical situations,
an intermediate node can only be partially trusted, as discussed {\sl
  e.g} in Ref.~\cite{solomons2022}.  All links are instead assumed to
be unconditionally secure.
To quantify this aspect we assume that
  every trusted node has a certain probability $p$ of being malicious
  and we define the security of a path, $s({\{a \to b\}})$ as the
  probability of finding only non-malicious trusted nodes along the
  path, {\sl i.e.}
\begin{equation}
\label{security}
s({\{a \to b\}}) \equiv (1-p)^{{\ell}_{\{a \to b\}}-1} 
%& = & \log(1-p) ({\ell}_{\{a \to b\}}-1)
\end{equation}
where ${\ell}_{\{a \to b\}}$ is the topological length of the path.
This definition yields $s=1$ for a path having topological length
$\ell=1$, i.e., no intermediate trusted nodes. Furthermore, it
correctly gives $s=0$ when $p=1$ and $s=1$ for $p=0$.

Within the model defined by Eqs.~\eqref{capacitance} and~\eqref{security},
 capacitance and security in general
compete, i.e., longer paths may have larger capacitance but at the price of
lower security.
To describe this trade-off we define the \change{communication efficiency} 
$\epsilon_{\alpha}(\{a \to b\})$ of a path
\begin{eqnarray}\label{efficiency}
  \epsilon_{\alpha}(\{a \to b\})&=&(1-\alpha) q({\{a \to
    b\}})+\alpha \log s ({\{a \to b \}}) \nn\\
%  &=& (1-\alpha) {\cal C}({\{a \to b\}})
%  + \alpha \log\lf(1-p\rg) ({\ell}_{\{a \to b\}}-1). \nonumber
\end{eqnarray}
where $\alpha \in [0,1]$ is a user-tunable control parameter
which gives more importance either to capacitance ($\alpha =0$) or to
security ($\alpha=1$).
For simplicity the \change{communication efficiency} is defined using the logarithm
of the security that is proportional to the path length.

Given the positions of users in space and assuming that a direct link
can be placed between any two of them, so that the network of all
possible links ${\cal G}_0$ is a fully connected (FC) graph, there is
an exponentially large number of possible paths between two nodes, $a$
and $b$.  For a given $\alpha$, an optimal path $\{a \to b\}^*$ is
defined as a path having maximal \change{communication efficiency} among all possible paths
\begin{equation}
\{a\to b\}^* = \argmax \{\epsilon_\alpha(\{a \to b\}) \}.
\end{equation}
Clearly,
in the limit $\alpha \to 0$, the optimal path maximizes
the capacitance, by going through many, physically close, intermediate nodes.
In the limit $\alpha \to 1$ the optimal path maximizes security and is thus
the direct link between the two nodes.
The global \change{communication efficiency} of a network ${\cal G}$ is defined as the
average, over all node pairs, of the \change{communication efficiency} of the optimal paths
defined over the network
\be
E_\alpha[{\cal G}]=\frac{1}{N(N-1)}\sum_{a,b}
\epsilon_{\alpha}(\{a \to b\}^*).\label{eff1}
\ee

It is important to remark that in general there can be more than a
single optimal path between two nodes.
As an example in Fig.~\ref{random-point}(a) we show two optimal
paths between the nodes $a$ and $b$ passing through the node $k$ and $k'$,
respectively. The two paths have the same \change{communication efficiency} since they have
the same topological length, $\ell=4$, and the same capacitance $Q=q(e_{cd})$,
being $e_{cd}$ the longest edge in both paths.

\subsection{Network optimization}

Given a set of $N$ users, ${\cal N}$,
%randomly
distributed in a
square of size $L$, and a tradeoff parameter $\alpha$, our goal is to
find a network connecting the $N$ users having maximal \change{communication efficiency} and
minimal number of links.  To reach this goal we start by determining
an optimal path for each pair of users.  This is achieved by means of
a novel algorithm described in detail in the Methods section.  Once an
optimal path is identified for each pair, we define the optimal
network, ${\cal G}^*_\alpha\equiv({\cal N},{\cal E}_\alpha^*)$, as the
graph union of the optimal paths, that is the network with node set
${\cal N}$ and edge set ${\cal E}_\alpha^*$ given by

\be
\label{gstar} {\cal E}_\alpha^*=\bigcup_{ab} \{a\to b\}^*.
\ee

This network has the maximum quantum \change{communication} efficiency, {\sl i.e.} no other
network ${\cal G'}\equiv({\cal N},{\cal E}')$ can have larger \change{communication efficiency}, {\sl i.e.},
\be
\label{optcond} E[{\cal G'}]\leq E[{\cal G}_\alpha^*]
\ee under the assumption of single-path~\cite{pirandola2019,solomons2022} routing.
For $\alpha = 0$ optimal
networks maximize the capacitance.  In this case, an efficient
optimization algorithm was proposed by Pollack~\cite{pollack1960}.  As
discussed in Ref.~\cite{hu1961}, beside maximizing network
capacitance, Pollack's algorithm minimizes the number of links
yielding as optimal network the maximum spanning tree (MST) connecting
the $N$ users~(Fig.~\ref{random-point}(b)).

The method developed in this work builds on Pollack's algorithm to
construct the optimal network $\cal G_\alpha^*$ as prescribed by
Eq.\eqref{gstar}.
Figure~\ref{random-point}(c-f) shows instances of $\cal G_\alpha^*$ for
different values of $\alpha$.  As one can see, it interpolates between
the fully connected network, realized for $\alpha= 1$, and a much
sparser network for $\alpha \to 0$.
We note that for $\alpha \to 0$,  $\cal G_\alpha^*$  does not necessarily reduce to the MST.
This is related to the fact that, as we explained above, the optimal paths are non-unique.
In principle, for generic $\alpha$ one could complement our
algorithm with further optimization techniques to reduce the number of links
of $\cal{G}_\alpha^*$ while preserving maximum \change{communication efficiency}.
This further development is deferred to future work.

\subsection{Maximal quantum \change{communication} efficiency networks}
\begin{figure*}
  \includegraphics[width=0.49\columnwidth]{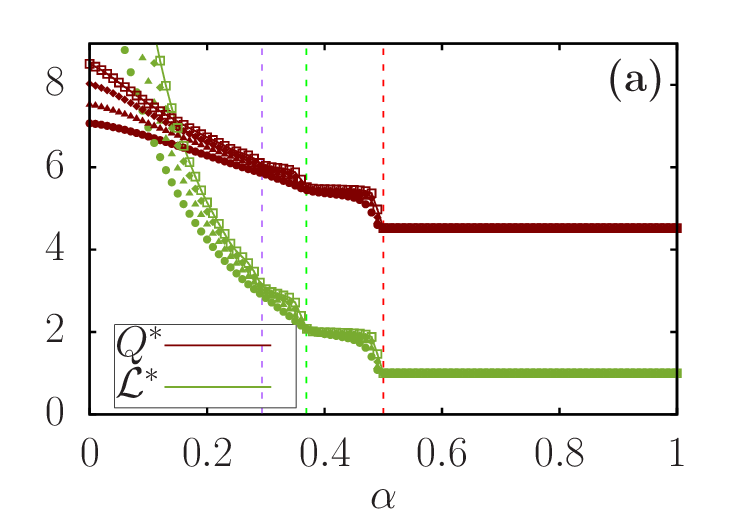}
  \includegraphics[width=0.49\columnwidth]{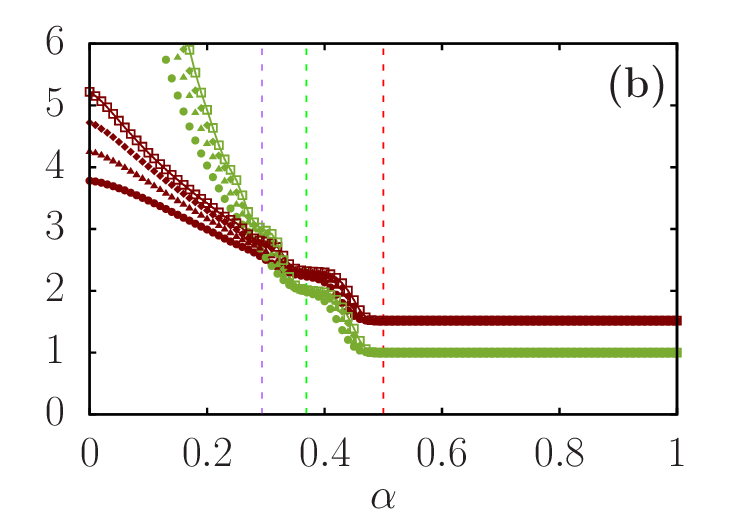}
  \includegraphics[width=0.49\columnwidth]{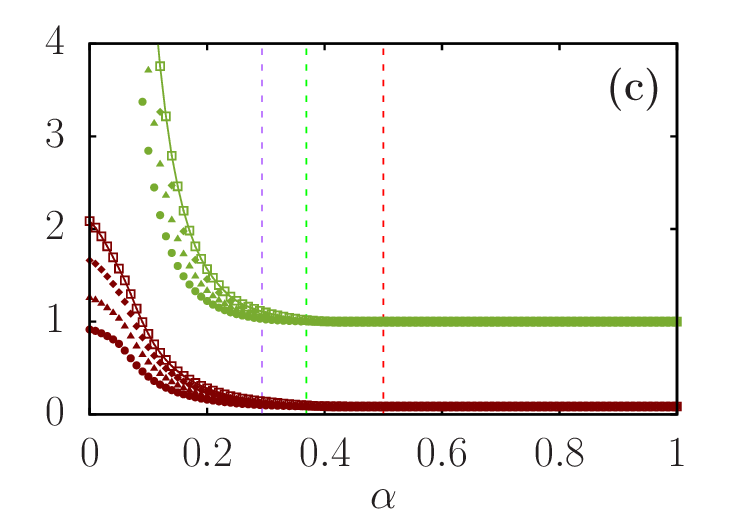}
 \includegraphics[width=0.49\columnwidth]{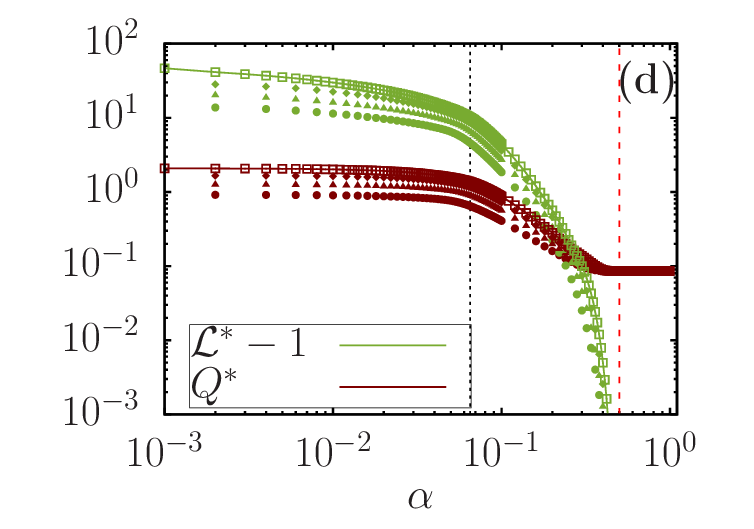}
 \caption{{\bf \change{Average properties of optimal networks.}}
   Average capacitance and path length,
    $\opt{Q}$ and $\opt{\mathcal{L}}$, on the \change{MQCE} networks as a
    function of $\alpha$ for $p=1-1/e\simeq 0.63$.  $N=256$ (circles),
    $N=512$ (triangles), $N=1024$ (diamonds), $N=2048$ (squares).  a)
    $L/\lambda_0 = 0.1$, b) $L/\lambda_0 = 1$, c) $L/\lambda_0 = 3$,
    d) $L/\lambda_0 = 10$.  Averages over $100$ realizations for
    $N=256$ and $N=512$, and over $10$ realizations for $N=1024$ and $N=2048$.
    Dashed lines correspond to the values of $\alpha_c^{0 \to 1}$
    (red), $\alpha_c^{1 \to 2}$ (green) and $\alpha_c^{2 \to 3}$
    (purple) evaluated using Eq.~\eqref{alphacm}.  }
  \label{transition}
\end{figure*}
For concreteness, in this subsection we assume a completely
random distribution of users in the square of size $L$.
As highlighted above, optimal networks maximize the quantum \change{communication} efficiency
functional defined by Eq.~\eqref{eff1}. Their structure and performance depend
sensitively on the user-defined parameter $\alpha$ and on the
distance-loss ratio, $L/\lambda_0$, that set the optimization regime.
To show how these {\it maximal quantum \change{communication} efficiency}
(\change{MQCE}) networks change across the different regimes, in
Fig.~\ref{transition} we plot their average capacitance,
\be
Q^*=\frac{1}{N(N-1)}\sum_{ab}q({ \{a\to b\}^*}), \,\,
\ee
and average topological length,
\be \opt{\mathcal{L}} =
\frac{1}{N(N-1)}\sum_{ab}{\ell}_{\{a \to b\}^*}
\ee
as a function of
$\alpha$ for different values of $L/\lambda_0$ ranging from the case
of weak losses ($L/\lambda_0=0.1$) to the case of strong losses
($L/\lambda_0 = 10$).

Let us consider first the case of weak losses, $L/\lambda_0 \ll 1$, 
%. In this case the analysis is easier because  VB easier rispetto a cosa?
where the distance between any pair of users is,
by construction, much smaller than the decay length $\lambda_0$;
in this case we distinguish three regimes.

(i) For values of $\alpha$ larger than a threshold value
$\alpha_c$ we observe ${\mathcal{L}^*}=1$, corresponding to all
optimal paths having $\ell_{\{a \to b\}^*}=1$.
In this regime ${\cal G}_\alpha^*$ coincides with the fully connected
network. Moreover, since the structure of ${\cal G}_\alpha^*$ does not
change with $\alpha$, the average capacitance is constant,
$Q^*=Q_{\text{FC}}$ (see Methods for its evaluation).
The critical value $\alpha_c$ is the lowest value of $\alpha$ such that,
for every pair of users $a,b$ in the system,
it is more efficient to connect them through the direct link
rather than using a trusted node $c$. 
This happens as long as the gain in \change{communication efficiency} due to
increased capacitance, is smaller than
the loss of \change{communication efficiency} associated to the
introduction of an intermediate node, equal to $\alpha \log(1-p)$,
yielding (see Methods for details)
\be
\label{alphac}
\alpha_c =[1-\log(1-p)]^{-1}.
\ee
(ii) On the left of this threshold value we observe the existence of a
``step'', i.e., an interval of $\alpha$ values over which
$\mathcal{L}^*$ is practically constant and equal to $2$. In this
interval all optimal paths have 1 intermediate node.  As $\alpha$
decreases other steps appear, corresponding to \change{MQCE} networks featuring
all optimal paths having topological length ${\cal L}^*=3$ (second step)
and ${\cal L}^*=4$ (third step). The steps become sharper in the large
$N$ limit where a simple calculation (see Methods) shows that paths
having $m$ intermediate nodes become more efficient than those with
$m-1$ intermediate ones for \be
%\alpha_c^{m-1 \rightarrow m}=\frac{\log(m/(1+m))}{\log(m/(1+m))+\log 2 \log(1-p)}.
\alpha_c^{m-1 \rightarrow m}=[1-\Delta(m) \log(1-p)]^{-1},
\label{alphacm}
\ee
where
\be
\Delta(m) = \frac{\log(2)}{\log(m+1)-\log(m)}.
\ee
Note that $\alpha_c^{0 \rightarrow 1}$ coincides with the $\alpha_c$ defined above.
The predictions $\alpha_c^{0 \rightarrow 1}=0.5$,
$\alpha_c^{1 \rightarrow 2} \approx 0.369$ and $\alpha_c^{2 \rightarrow 3} \approx 0.293$
are in very good agreement with the positions of the steps for $p=1-1/e$, appearing
in Fig.~\ref{transition}(a).

(iii) As $\alpha$ tends to 0, \change{MQCE} networks increasingly resemble the
maximum spanning tree. In this limit the average topological length of
optimal paths tends to increase as a power-law with $N$ (see Methods).
Let us now consider the opposite case of strong losses, $L \gg \lambda_0$.  In this
limit the phenomenology is different and more difficult to interpret because,
while most distances are much larger than $\lambda_0$, still some
pairs of users are at a distance smaller than the decay length.  These
pairs are responsible for the observation that $\opt{\mathcal{L}}>1$
as soon as $\alpha$ gets smaller than the threshold value still given
by Eq.~\eqref{alphac} (see Fig.~\ref{transition}(c-d)).  At variance
with the previous case, for $\alpha<\alpha_c$ the average topological
length of the optimal paths $\opt{\mathcal{L}}$ grows rapidly as
$\alpha$ is reduced, showing no steps at integer values.  Moreover, there
exist a value $\overline{\alpha}$ such that
for $\alpha<\overline{\alpha}$, $\opt{\mathcal{L}}$ and $Q^*$ vary
much less and assume values close to those of the MST network (see Methods).

\begin{figure*}[t]
\includegraphics[width=0.65\columnwidth]{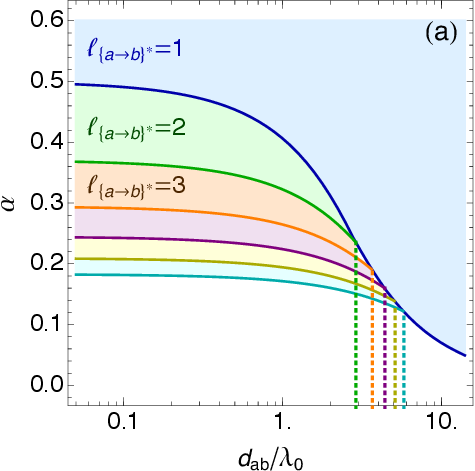}
\includegraphics[width=0.65\columnwidth]{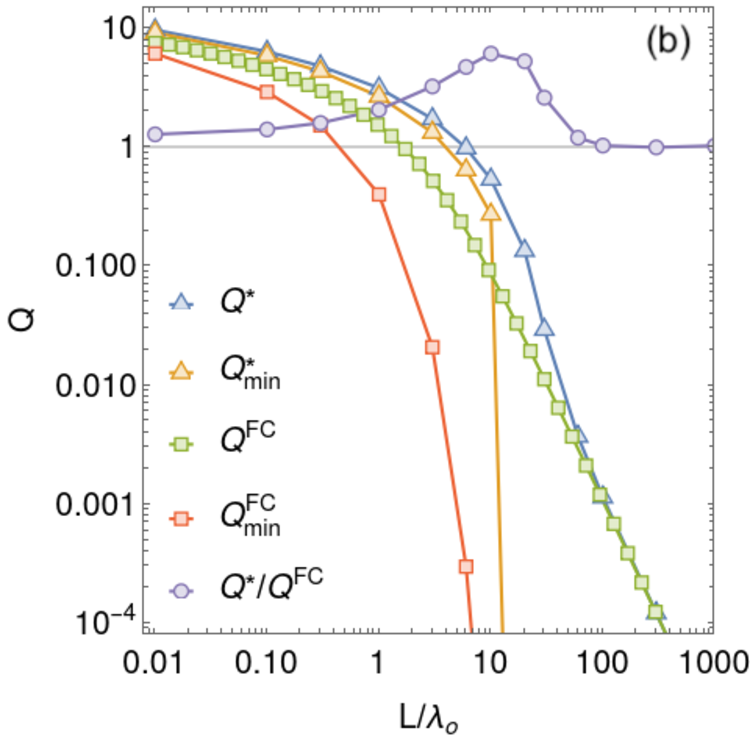}
\includegraphics[width=0.65\columnwidth]{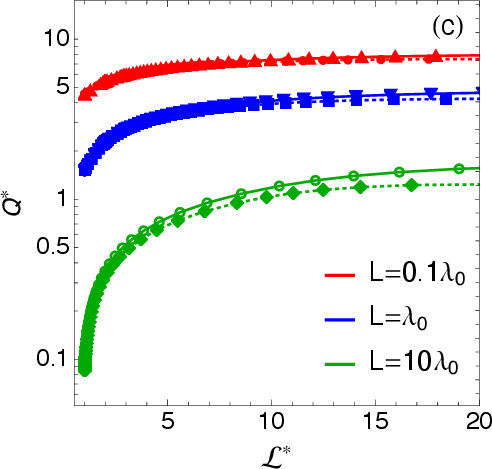}
\caption{{\bf \change{The phase diagram and the behavior of the average capacitance.}}
  (a) Transitions between the different regimes for $p=1-1/e$
    and $N \to \infty$, representing the number of intermediate nodes
    in the optimal path for a pair of nodes at distance $d_{ab}$ as a
    function of $d_{ab}$ and $\alpha$.  Note that the curve separating
    the ``direct link region'' from the other regions has slope
    discontinuities corresponding to the dashed vertical lines.
    (b) Behavior of the average capacitance $Q^*$, and of the minimum
    capacitance $Q_{\rm min}$ as a function of $L/\lambda_0$ 
    with $\alpha=0.1$, $p=0.1$, $N=100$ for the \change{MQCE} network \change{,} the
    FC network (computed analytically, see Methods).
    (c) Plot of the average capacitance of the \change{MQCE} network as a function
    of the average topological length for $p=1-1/e$, $N=512$ (dotted lines),
    $N=1024$ (solid lines) and various values of $L/\lambda_0$.}
  \label{QvsL}
\end{figure*}

A partially quantitative understanding
of the nature of the different regimes
can be achieved by considering how the properties of the optimal path
linking a generic pair of users, $a$ and $b$, change, as a function of
their distance $d_{ab}$ and of $\alpha$, in the limit $N \to \infty$.
In Fig.~\ref{QvsL}(a) we highlight with different colors
the regions corresponding to different topological lengths of the
optimal path between $a$ and $b$.
Boundaries shown in Fig.~\ref{QvsL}(a) are calculated analytically
(see Methods).
 For large values of $\alpha$ security dominates and
 the optimal path is, for any $d_{ab}/\lambda_0$, the direct
 connection.  When $\alpha$ is reduced the optimal path has a
 different behavior depending on the ratio $d_{ab}/\lambda_0$.
  For small $d_{ab}/\lambda_0$ , when
 $\alpha$ is reduced it becomes more efficient to go through indirect
 paths going through $m = 1, 2, 3,\ldots$, intermediate trusted nodes
 and so on, which are equally spaced along the line connecting $a$ and
 $b$.
 For larger $d_{ab}/\lambda_0$, the gain in capacitance provided
 by a single intermediate trusted node is not sufficient to compensate
 the loss in security and, as $\alpha$ is reduced, the first
 transition occurs between the direct link and a path going through 
 $\overline{m}>1$ trusted nodes.  This first transition is followed, as in
 the previous case, by a one by one increase in the number of
 intermediate nodes as smaller $\alpha$ values are considered. For
 $d_{ab}/\lambda_0 \gg 1$, $\overline{m}$ grows linearly with $d$ and
 the position of the first transition scales as $1/d$.

The phenomenology for the whole network shown in Fig.~\ref{transition}
results from the superposition of the behavior just depicted for all
user pairs whose distances are distributed according to Eq.~\eqref{gv}
(see Methods).  Specifically, if $L \ll \lambda_0$ all pairs belong to
the small distances regime of Fig.~\ref{QvsL}(a). As a consequence
they all undergo the same transitions for the same values of $\alpha$,
thus generating the steps observed in Fig.~\ref{transition} (a).  For
$L \gg \lambda_0$ instead, most pairs are at distances larger than the
decay length, but a few are still at distances much smaller than
$\lambda_0$.
%The latter are responsible for the fact that ${\mathcal{L}}>0$ as soon as $\alpha<\alpha_c$.
%The former, which constitute the bulk of the system, fall in the right part of
%Fig.~\ref{PhaseDiagram}.
Therefore, depending on the exact value of $d_{ab}/\lambda_0$, each
user pair undergoes different transitions for different values of
$\alpha$.
All these transitions get ``mixed'', thus explaining the lack
of steps in this case and a smooth growth of the observables between
$\alpha_c$ and $\overline \alpha$ (see Fig.~\ref{transition} (d)).

\subsubsection{Performance of \change{MQCE} networks}
By balancing security and quantum capacitance, \change{MQCE} networks yield the
optimal strategy to connect $N$ users for quantum communications and,
at the same time, they represent a customizable benchmarking tool for
quantum communication networks.
To illustrate the performance of \change{MQCE} networks we start by comparing
their quantum capacitance to that of FC networks.  As one can see
in Fig.~\ref{QvsL}(b), the outcome depends sensitively on the ratio
$L/\lambda_0$.  For $L/\lambda_0 \ll 1$ a limited increase of the
average capacitance with respect to the FC network is observed. For
values of $L$ of the order of $\lambda_0$ the improved performance of
the \change{MQCE} network becomes more significant.  In this range of
$L/\lambda_0$ values, the most important improvement concerns the
minimum capacitance.  While for FC networks some links have a strongly
degraded capacitance, several orders of magnitude smaller than the
average $Q^*$, in the optimized network $\min Q$ is only slightly
smaller than the average, thus guaranteeing that communication is
possible among any pairs of users.  For very large values of
$L/\lambda_0$ the \change{MQCE} network tends to coincide with the FC one, as
the weight of capacitance in the quantum \change{communication} efficiency functional becomes
extremely small.  In this regime the average capacitance $Q^*$ is
essentially the same in the two networks and decays as $(L/\lambda_0)^{-2}$.
%, but they are
%so low that this limit is of no practical interest.
In Fig.~\ref{QvsL}(b) we also note that in \change{MQCE} networks the minimum
capacity reaches the threshold of one target bit per use of the channel,
$Q_{\rm min}=1$, for values of $L/\lambda_0$ about one order of
magnitude larger than in standard FC networks. For fixed values of
$\lambda_0$, this implies that \change{MQCE} networks can provide the minimum
quantum communication rate standard across much wider regions of
space.  Specifically, for the parameters in Fig.~\ref{QvsL}(b),
the maximum value of $L$ to have at least 1
target bit transferred between any user of the network goes from
$\approx 0.5 \lambda_0$ (in the FC network) to $\approx 5 \lambda_0$
in the \change{MQCE} network.
A natural question arises concerning
how the the security requirement affects the maximal achievable rate.
To clarify this point in Fig.~\ref{QvsL}(c) we plot the average
capacitance of \change{MQCE} networks, $Q^*$ as a function of the average
topological length $\opt{{\cal L}}$, for different values of the ratio
$L/\lambda_0$.
It turns out that it is sufficient to allow $\opt{\cal L}$ to grow from 1
(FC network) to 3 or 4 to ensure a considerable increase in the average
capacitance, even one order of magnitude for large $L/\lambda_0$.

\subsubsection{Structure of \change{MQCE} networks}

\begin{figure*}[t]
  \includegraphics[width=\textwidth]{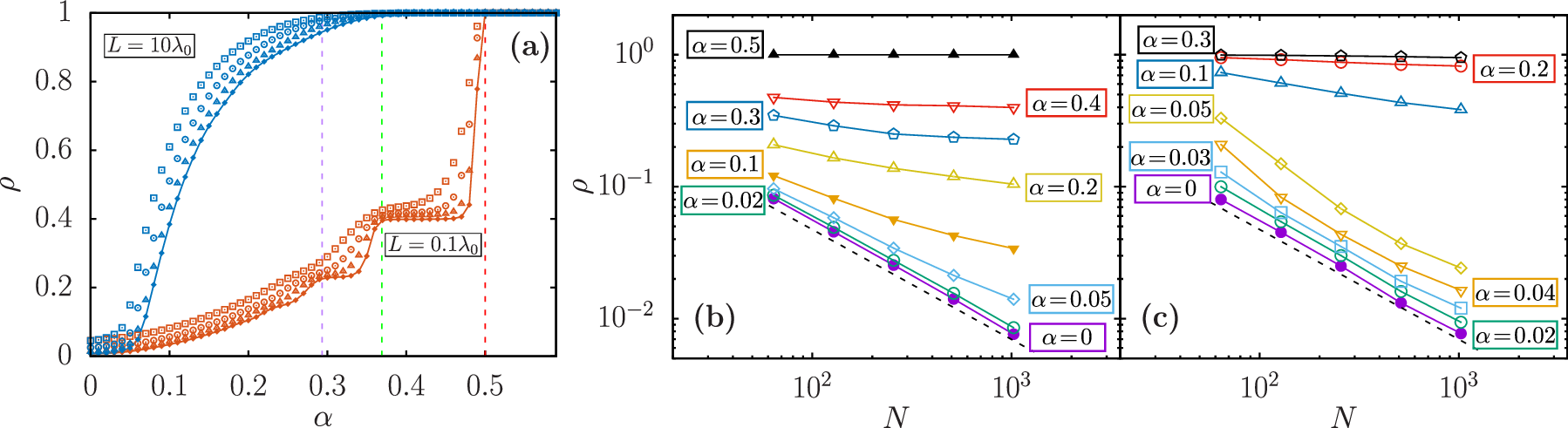}
  \caption{{\bf \change{
    The link density of the optimal networks.}}
    (a) Density $\rho$ as a function of $\alpha$, for $p=1-1/e$
    and two values of $L/\lambda_0$
    Averages over $100$ realizations for $N=128~ (\text{squares}),
    N=256~(\text{circles})$, and over $10$ realizations
    for $N=512~(\text{triangles}), N=1024~(\text{diamonds})$.
    Dashed lines correspond to the values of $\alpha_c^{0 \to 1}$ (red),
    $\alpha_c^{1 \to 2}$ (green) and $\alpha_c^{2 \to 3}$ (purple)
    evaluated using Eq.~\eqref{alphacm}.
    (b) Dependence of the density on $N$ for $L/\lambda_0 = 0.1$.
  (c) Dependence of the density on $N$ for $L/\lambda_0 = 10$.}
  \label{topological}
\end{figure*}

It is interesting also to analyze the structural features of the
networks that the optimization algorithm generates.
In Fig.~\ref{topological} we report
the dependence on $\alpha$ of the link density
of the optimized network, $\rho=\mean{k}/(N-1)$,
which is correlated with the cost to build the infrastructure.
In all cases the density interpolates between a maximally
dense network, the FC, (for $\alpha>\alpha_c$)
and a much sparser network (for $\alpha \to 0$).
As mentioned above, our algorithm does not exactly reproduce,
in the limit $\alpha \to 0$, the MST (which has density $2/N$).
The scaling of $\rho$ vs $N$, which exhibits
a decay $N^{-\omega}$, with an effective exponent
to $\omega \approx 0.83$ implies that the optimized network
has an average degree growing sublinearly with $N$.
For large values of $\alpha$ the densities tend to converge to 
a finite limit, indicating that the networks are dense.
For small but finite $\alpha$ values the initial decay with $N$
appears analogous to the $\alpha=0$ case, 
there is some evidence that for any $\alpha>0$ the density
tends to a constant.

\begin{figure}
  \includegraphics[width=\columnwidth]{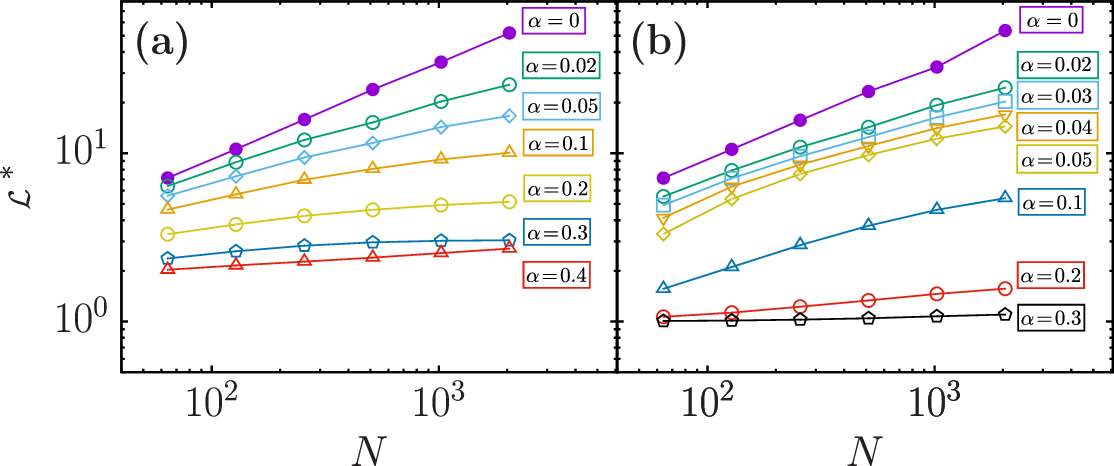}
  \caption{{\bf \change{Size dependence of the average length of optimal paths.}}
    Plot of $\mathcal{L}^*$ vs $N$ for several values of $\alpha$,
%    a) $p=0.1$, $L/\lambda_0=0.1$;
%    b) $p=0.1$, $L/\lambda_0=10$;
    $p=1-1/e$, (a) $L/\lambda_0=0.1$, (b) $L/\lambda_0=10$.
%    The dashed line is the scaling $N^{5/8}$ expected for the MST.
}
    \label{SvsN}
\end{figure}

Additional information on the optimized network
is provided by Fig.~\ref{SvsN}, where the average length of optimal
paths $\mathcal{L}^*$ is plotted as a function of the number of users.
For small values of $L/\lambda_0$ and relatively large values
of $\alpha<\alpha_c$ it is clear that $\mathcal{L}^*$ goes to a constant
in the large-$N$ limit (see Fig.~\ref{SvsN}(a)).
This mirrors the presence of steps in Fig.~\ref{transition}.
For smaller values of $\alpha$ and larger values of $L/\lambda_0$
the average optimal path length exhibits an initial power-law growth
with $N$ followed by a smooth crossover to a constant value.
%For $\alpha \to 0$, we recover the power-law growth $N^{5/8}$
%expected for the MST (see Methods),
%(consistently with the fact that the MST is underlying the
%optimized network).

%\subsection{The betweenness distribution}

\begin{figure}
  \includegraphics[width=0.49\columnwidth]{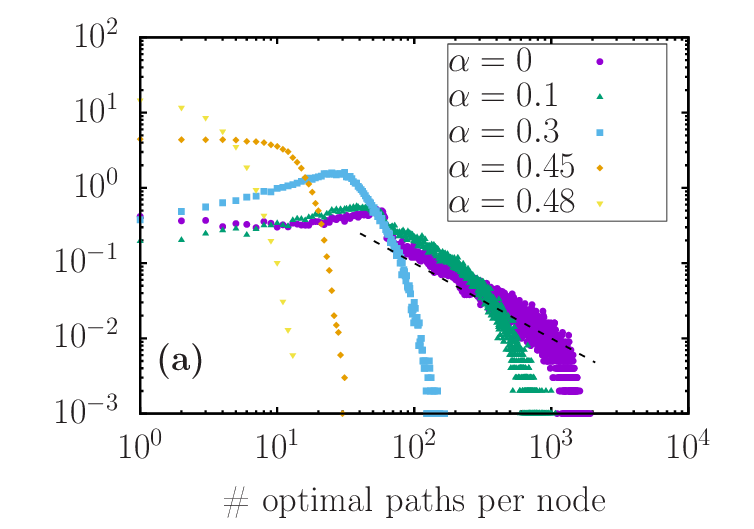}
  \includegraphics[width=0.49\columnwidth]{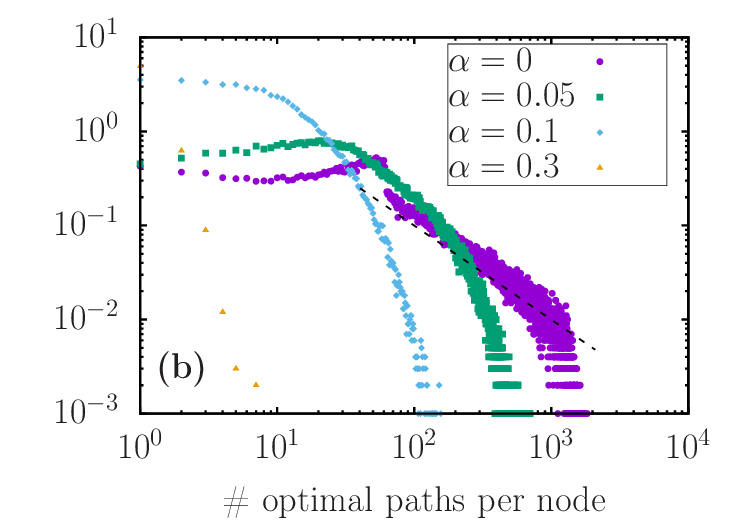}
  \caption{
    {\bf \change{The impact of individual nodes on security is heterogeneous.}}
Distribution of the number of nodes having a given modified betweenness for
$N=64$, $p=1-1/e$, $L/\lambda_0=0.1$ (a), $L/\lambda_0=10$ (b),
and several values of $\alpha$.\change{ The lower $\alpha$, the broader the distribution becomes. Black dashed lines are guides to the eye.}}
\label{bet}
\end{figure}

The optimal network can be also characterized
by measuring a quantity analogous to the betweenness usually considered in network
analysis. The betweenness~\cite{Newmanbook} of a node is the number of shortest paths
among all pairs of nodes in the network that go through that node.
For our purposes it is useful to define a modified betweenness where,
instead of topological shortest paths, optimal paths are considered.
Such a quantity provides a measure of the relevance of users, i.e.,
how crucial is their presence (and how damaging their removal).
\change{Nodes with high betweenness have, just because of their position in
the topology, a high impact on the security of the network.}
In Fig.~\ref{bet} we plot the histogram of the number of nodes having a given
modified betweenness for various $\alpha$.
For $\alpha>\alpha_c$ the optimal network is fully connected and the values
of the betweenness are all zero.
When $\alpha$ becomes smaller than $\alpha_c$ a homogeneous distribution
appears. As $\alpha$ is progressively reduced the distribution gets more
heterogeneous, becoming extremely broad for $\alpha \to 0$.
In such a case some nodes are particularly crucial and the network is
overall highly vulnerable to external attacks.

\vspace{3cm}

\section{Methods}
\subsection{Optimization algorithm}
The task of constructing \change{MQCE} networks cannot be addressed using
dynamic programming methods such as Dijkstra's~\cite{dijkstra1959}, Prim's~\cite{prim1957} or Pollack's~\cite{pollack1960}   algorithms. Specifically,  because of the inherently nonlocal nature of the
problem arising from the interplay of  topological and geometric terms in the \change{communication efficiency} functional,  if the optimal path between $a$ and $b$ goes through node $c$ nothing
guarantees that the subpath between $a$ and $c$,  belonging to $\lf\{a\rightarrow b\rg\}^*$, yields also the optimal path
between $a$ and $c$. Consequently, we have to resort to alternative approaches to tackle this problem.
In Ref.~\cite{pollack1960} Pollack presents, among others, a matrix-based
algorithm
for solving the maximum capacity route problem, which is equivalent for us
to finding the maximum capacitance between any pair of users. Starting
from a matrix 
whose element $Q^{(0)}_{jk}$ is the capacitance of the link between
node $j$ and $k$ given by Eq.~\eqref{chq} and $Q^{(0)}_{jj}=\infty$,
an iterative procedure is defined. The elements of $Q^{(m)}$ at step
$m$ are defined by \be Q^{(m)}_{jk} = \max_\ell \left[ \min(Q^{(0)}_{j
    \ell},Q^{(m-1)}_{\ell k}) \right] \ee where $\ell=1,\ldots,N$.  In
this way the element $Q^{(m)}_{jk}$ is the maximum
capacitance of paths between $j$ and $k$ going through at most $m$
intermediate nodes.  Iterating, the process, convergence is reached at
most when $m+1=N-1$.  Individual elements of $Q^{(m)}$ provide at convergence
the maximum capacitance between all node pairs.
We modify Pollack's original algorithm as follows.
Given $\alpha$, at each iteration $m$ we evaluate the quantity
\change{
\be
\epsilon_{jk} = (1-\alpha) Q^{(m)}_{jk}+\alpha m \log(1-p).
\label{eq15}
\ee
}
This is the maximal \change{communication efficiency} of paths between $j$ and $k$ of 
length at most $m+1$.
\change{As $m$ is increased, the first term in Eq.~\eqref{eq15} grows and tends
to a constant, while the second is negative and is linear in $m$. Hence, 
for each pair $(j, k)$, $\epsilon_{jk}$ reaches a maximum as a function of $m$
for a value $m^*_{jk}$.}
Note that $m^*$ needs not to be the same
for all pairs.
The \change{communication efficiency} of the optimal path between $j$ and $k$ is thus

\be
\epsilon(\{j \to k\}^*) = (1-\alpha) Q^{(m^*_{jk})}_{jk}+\alpha \log(1-p) m^*_{jk}.
\ee

This procedure gives the \change{communication efficiency} of the optimal paths between any two users.
In order to construct one of these (in principle many) paths we proceed
as follows.
The capacitance $Q^{(m^*_{jk})}_{jk}$ associated to the optimal path between $j$
and $k$
necessarily appears in the initial matrix $Q^{(0)}$, as element $Q^{(0)}_{ln}$.
All entries of $Q$ are assumed to be different.
This indicates that the optimal path between $j$ and $k$ necessarily goes
through the link between
$l$ and $n$, which, indeed, determines the capacitance of the optimal path.
Starting from the matrix $Q^{(0)}$ we construct the unweighted graph $A$
containing an edge between all node pairs such that the corresponding element
in $Q^{(0)}$ is larger than  $Q^{(0)}_{ln}$. In this graph $A$ we find the
optimal path as the topological shortest path going from $j$ to $k$ with the
constraint that it goes through the link $(l,n)$.
Clearly, although the link between $l$ and $n$ is uniquely determined
there are often many alternative shortest paths going through it;
this gives rise to a large number of degenerate optimal paths.

\subsection{The efficiency of the fully connected network}

To calculate the efficiency of a fully connected network, which
coincides with its capacitance, we need
the probability distribution for the distance between any pair of points
in a square of size $L$.
This quantity coincides with the distribution of distances
between two points randomly distributed in the square, which
is a special case of the distribution for a generic
rectangular substrate derived in Ref.~\cite{philip2007probability}.
Note that the quantity $N$ does not play any role.
Considering nodes distributed in a square of side $L$,
the distance distribution is
\be
g(d) = \frac{1}{L} f\left(\frac{d}{L} \right)
\ee
where
\begin{widetext}
\be
f(z) = 2z \left\{
  \begin{array}{lll}
    \pi-4z+z^2 & & 0 \le z \le 1 \\ \newline
    4 \arcsin\left(1/z\right)-(2+\pi)+4\sqrt{z^2-1}-z^2 & & 1 \le z \le \sqrt{2}.
  \end{array}
  \right.
  \label{gv}
\ee
\end{widetext}
%and is represented in Fig.~\ref{Philip}.
%\begin{figure}
%  \includegraphics[width=0.9\columnwidth]{P_Distr.eps}
%  \caption{Plot of $f(z)$ (see Eq.~\eqref{gv}).}
%  \label{Philip}
%\end{figure}
Given this distribution, the value of the capacitance averaged over all node pairs is then:
\be
Q_{\text{FC}} = - \frac{1}{\ln{2}} \int_0^\infty dz ~f(z) \ln(1-e^{-z/{\tilde \lambda}}),
\label{AQ}
\ee
where ${\tilde \lambda}=\lambda_0/L$.
%Notice that $Q$ quantity depends only on $L/\lambda_0$, with no dependence
%on the number of points $N$ \NoteLC{questo perche' e' una densita' ricavata nel limite di infiniti punti, giusto? Non dipende da $N$ perche' $N\to \infty$}.
In the limit $L \ll \lambda_0$ one can
safely approximate $e^{-z/{\tilde \lambda}} \approx 1-z/{\tilde \lambda}$, hence
\be
Q_{\text{FC}} = - \frac{\langle \ln(z/\tilde{\lambda} \rangle)}{\ln{2}}.
\ee

In the limit $L \gg \lambda_0$ instead, taking
\be
\ln(1-e^{-d/\lambda_0}) \approx - e^{-z/\tilde{\lambda}},
\ee
we can write
\be
Q_{\text{FC}} \approx \frac{1}{\ln{2}} \int_0^\infty dz f(z) e^{-z/\tilde{\lambda}}.
\ee
Since for small $z$ we have $f(z) \approx 2 \pi z$, then
\be
Q_{\text{FC}} \approx \frac{2 \pi}{\ln{2}} \left(\frac{\lambda_0}{L} \right)^2.
\label{AQexp}
\ee
In this regime, the value of $Q_{\text{FC}}$ is determined by
the contribution of the few links shorter than $\lambda_0$, which have
a large capacitance.

\subsection{Optimization regimes for a single pair}

We consider $N \to \infty$.
In this limit we can always assume that for any pair of users
at distance $d_{ab}$, there are $m$ ($\forall m$) equally spaced
intermediate users along the line
joining $a$ and $b$, so that the distance between nearest neighbouring
nodes is $d_{ab}/(m+1)$.
The efficiency of the path going through $m$ intermediate nodes
(including the direct link, which is the case $m=0$) is therefore:
\be
E(m) = -(1-\alpha)\log_2(1-e^{-d_{ab}/[(m+1)\lambda_0]})+\alpha \log(1-p) m.
\label{Em}
\ee
For sufficiently large $\alpha$ it is more efficient to use the direct link, $m=0$.
As $\alpha$ is reduced it becomes more efficient to use indirect paths, with $m>0$.

For $d_{ab} < \lambda_0$, if one plots $E(m)$ vs $\alpha$ for
$m=0,1,2,3,\ldots$ one observes that the intersections of the lines for $m>0$
with the line for the direct link ($m=0$) occur for decreasing $\alpha$
as $m$ is increased.
As a consequence, starting from $\alpha=1$, at a
given $\alpha_c^{0 \rightarrow 1}\equiv \alpha_c$
it becomes more efficient to follow the path of length 2 (i.e.,
with $m=1$ intermediate nodes) rather than the direct link (with $m=0$).
Analogously, for a smaller $\alpha_c^{1 \rightarrow 2}$ the path of length 3
becomes more efficient than the path of length 2 and so on.
Hence we observe a series of transitions, 
where it becomes more efficient to use paths with $m$ intermediate nodes
with respect to paths with $m-1$ nodes, with increasing $m=1,2,3,\ldots$
The location $\alpha_c^{m-1 \rightarrow m}$ of the $m$-th transition
is given by the condition
\be
E(m)=E(m-1),
\label{Conditionm}
\ee
which, in the limit $d_{ab} \to 0$, yields Eq.~\eqref{alphacm}.
These transitions are indicated by the solid lines in Fig.~\ref{QvsL}(a).
For $d_{ab} \gg \lambda_0$ instead,
the intersection of $E(m)$ with $E(0)$ (efficiency of the direct link)
occurs at a value of $\alpha$ that grows initially with $m$, up to a
value $\overline{m}$.
After this value the position of the intersection decreases with $m$.
%We call $\overline{\alpha}$ the value of the intersection corresponding
%to $\overline{m}$.~\NoteCC{Questo $\overline \alpha$ non \`e lo stesso
%  definito nel main}
This means that, instead of having a transition from the direct
link to a path with one trusted node, for
$\alpha=\alpha^{0 \rightarrow \overline m}$ there is a transition from
the direct link to a path going through $\overline m>1$ intermediate nodes.
This first transition is followed by other transitions of the same
kind as before, in which the number of intermediate nodes is increased by
1, $\alpha_c^{m-1 \rightarrow m}$ for $m > \overline{m} +1$.
The value of $\overline{m}$ is found by determining $\alpha_c^{0 \rightarrow m}$
using the condition
\be
E(m)=E(0),
\label{Conditionm2}
\ee
and checking, for a given $d_{ab}/\lambda_0$, which of the
$\alpha_c^{0 \rightarrow m}$ is the largest.
In the limit of large $d_{ab}/\lambda_0$, $\overline m$ gets large as well;
in other words, starting from $\alpha=1$ one jumps from the direct link to
a path of larger and larger topological length, but this happens for
smaller and smaller values of $\alpha_c^{0 \rightarrow \overline{m}}$,
vanishing as $1/d_{ab}$.
Fig.~\ref{QvsL}(a) represents the scenario just described.

\subsection{Optimization regimes for the whole network}

The previous discussion (and Fig.~\ref{QvsL}(a)) consider a single
pair, with given distance $d_{ab}$, and is exact in the limit $N \to
\infty$.  The phenomenology of the whole system of size $L$ is the
superposition of what happens for each pair in the system where
distances, spanning the range $0<d_{ab}<\sqrt{2}L$ are distributed
according to Eq.~\eqref{gv}.  For $L/\lambda_0 \to 0$ all distances
$d_{ab}$ are necessarily much smaller than $\lambda_0$.  In the
expression for $E(m)$ one can expand the exponential to first order
and the values $\alpha_c^{m-1 \to m}$ obtained by solving the equation
$E(m)=E(m-1)$ do not depend on $d_{ab}$.  Hence the transitions for
all pairs occur exactly for the same values of $\alpha$, given by
Eq.~\eqref{alphacm}.  Thus the average optimal path length
$\mathcal{L}$ exhibits, as a function of $\alpha$, steps which are, in
the limit $L/\lambda_0 \to 0$, perfectly sharp
(Fig.~\ref{transition}(a)).

For generic $L/\lambda_0$, distances $d_{ab}$ are distributed over the
range between $0$ and $\sqrt{2}L$.
For the smallest of them the scenario just depicted applies and
transitions occur for the values written in Eq.~\eqref{alphacm}.
But for any finite $L \ll \lambda_0$ there are corrections to the
values of Eq.~\eqref{alphacm}, shown by the decreasing behavior of the
curves in Fig.~\ref{QvsL}(a)).
This explains why, for $0< L \ll \lambda_0$, steps are not perfectly sharp
but broadened (even for $N \to \infty$): the transitions from $m-1 \to m$ intermediate
nodes occur, for the various pairs, at slightly different values of $\alpha$.
As $m$ grows the transitions are closer (see the denser lines in Fig.~\ref{QvsL}(a));
the broadening becomes stronger so that steps of high order $m$ cannot be clearly identified.

For $L/\lambda_0$ not infinitesimally small but still smaller than 1,
an estimate of the average position of the steps is obtained by using
the condition~\eqref{Conditionm}, after setting
$d_{ab}=\mean{d_{ab}} \approx 0.52 L$~\cite{philip2007probability}.
The values obtained, which are a decreasing function of $L$.
explain why the positions of the observable steps in Fig.~\ref{transition}(b-c),
which for $L/\lambda_0 \to 0$ are given by Eq.~\eqref{alphacm},
decrease as $L$ grows.
At the same time the increase of $L$ leads to an increased ``mixing''
of the transitions. As a consequence, the steps get less sharp
and the concept of step progressively loses meaning.

For large $L/\lambda_0$, the connection between the regimes
for a single pair and the overall behavior of the system is more involved.
Indeed, if $L/\lambda_0 \gg 1$ despite the fact that
for the overwhelming majority of pairs $d_{ab}/\lambda_0 \gg 1$
there are always some pairs of users whose distance is $d_{ab} \ll \lambda_0$.
For them, the optimal path involves a number $m$ of intermediate nodes
growing one at a time for the values of $\alpha$ given by Eq.~\eqref{alphacm}.
%(see the left part of Fig.~\ref{QvsL}(a)).
This explains why, for any $L$, one observes ${\mathcal{L}}>0$ as soon as
$\alpha<\alpha_c=[1-\log(1-p)]^{-1}$ (see Fig.~\ref{transition}(d)).
However, for $L \gg \lambda_0$ the overwhelming majority of
distances are much larger than $\lambda_0$ and the length of the optimal
path is  described by the right part of Fig.~\ref{QvsL}(a).
For a given $\alpha<\alpha_c$ the optimal path for these pairs involves
$m$ intermediate nodes, with $m$ assuming a range of values
depending on the precise value of $d_{ab}/\lambda_0$.
For this reason the average length of optimal paths $\mathcal{L}$
does not assume integer values but instead it changes continuously
with $\alpha$, thus explaining the lack of steps in this case.
This “mixing” involves all distances up to the largest one
in the system, $d_{max} = \sqrt{2}L$, for which the transition to
optimal paths longer than $1$ occurs for the smallest $\alpha$ value.
We can therefore estimate that for
$\alpha < \alpha_c^{0 \rightarrow \overline{m}}(d_{max})$
all pairs are connected by topologically long paths and
the network is essentially a MST.
This leads to the identification of the threshold $\overline \alpha$ observed
in Fig.~\ref{transition}(d) as
\be
\overline \alpha \approx \alpha_c^{0 \rightarrow \overline{m}}(d=\sqrt{2}L)
\sim 1/L.
\ee
\subsection{Capacitance of the Maximum Spanning Tree}

The length of the longest link (i.e. the link with lowest capacitance)
in an Euclidean MST is given by~\cite{Penrose1997}
\[\max d_{e} \simeq L\sqrt{\frac{\log N}{\pi N}}, ~~~~~~N \gg 1.\]

Assuming that the majority of the optimal paths pass through the link with
minimum capacitance, we can estimate $Q_{\text{MST}}$ as the capacitance of such a link.
Hence we have
\begin{equation}
  Q_{\text{MST}} \simeq -\log_2
  \left[1-\exp\left({\frac{L}{\lambda_0}\sqrt{\frac{\log N}{\pi N}}}\right) \right],
\end{equation}
which is in good agreement with the results in
Fig.~\ref{transition} for $L \ll \lambda_0$ and is increasingly
accurate as $N$ grows also in the other cases.

\section{Conclusions}
The realization of large-scale quantum communication networks is a
task of crucial relevance for quantum cryptography and quantum
computing applications. Existing quantum network implementations
employ intermediate trusted nodes as a practical and efficient means
to connect remote users, thereby overcoming the limitations imposed by
rate/distance bounds. Relying on trusted nodes, however, carries the
inherent risk associated with the probability of encountering
malicious nodes. Given these constraints, in this work we addressed
the problem of designing optimal quantum communication networks
connecting a set of users randomly distributed in a square of size
$L$. For each pair of users, we determined the optimal path connecting
them by maximizing the quantum \change{communication} efficiency of the path and we
constructed the optimal network, called maximal quantum \change{communication} efficiency
network as the graph union of the optimal paths. \change{MQCE} networks
therefore provide the optimal balance between security and quantum
communication rate and they interpolate between Maximum Spanning Trees
and Fully Connected networks, representing, respectively, the topologies
having maximum quantum capacitance and maximum security.
We carefully analyzed the performance of the
\change{MQCE} networks showing that the optimization can largely 
increase the average capacitance while keeping high levels of security.
We also analyzed structural properties of \change{MQCE} networks 
by means of numerical and analytical methods, showing 
how the tradeoff between capacitance and security
affects their topological properties.
\change{Our work proposes a systematic and scalable approach for quantum communication network optimization that relies on the construction of a network model and the corresponding quantum communication efficiency functional. So far we considered simple networks featuring only trusted nodes and lossy bosonic links but our work lays the basis for the study of general quantum communication networks.}

%\NoteCC{We promised in the rebuttal to add a concise take-home message, but I
%do not know what to write.}
%
We expect the performance of \change{MQCE} networks to be significantly affected by the
spatial distribution of points; for simplicity we
assumed a uniform distribution but it would be interesting to consider
other possibilities.  \change{A further} crucial assumption underpinning our work
is that of single-path routing; more powerful routing strategies,
where systems are transmitted in parallel through different quantum
communication channels have been proposed to improve the
capacitance~\cite{pirandola2019,das2021} or the security~\cite{solomons2022}.
In these situations, an extension of our algorithm may provide a way to combine
the two approaches to fullfill simultaneously well-defined capacitance
and security requirements. The degeneracy of the optimal path could
then acquire further practical relevance.
\change{Eventually, our method could be extended to  different quantum key distribution schemes such as entanglement-based quantum key distribution \cite{bennett1993,long2002,zhang2017}.}
In conclusion, we stress that our freely available code~\cite{Code}
can be straightforwardly used to design optimal quantum communication networks
for given location of users.

\section*{Acknowledgements}
This work was funded by the HEISINGBERG “Spatial Quantum Optical Annealer for Spin Hamiltonians” EU Research and Innovation Project under Grant Agreement No.~101114978. V.B. acknowledge support from Project PNRR MUR PE$\_0000023$-NQSTI ﬁnanced by the European Union Next
Generation EU. V.B. and L.P. acknowledge support from PON Ricerca e
Innovazione 2014-2020 FESR /FSC - Project ARS01$\_00734$ QUANCOM,
Ministero dell’Università e Ricerca and PNRR MUR project CN$\_00000013$-ICSC ﬁnanced by the European Union Next Generation EU. The funders
played no role in study design, data collection, analysis and interpretation of
data, or the writing of this manuscript.

\section*{Competing interests}
All authors declare no financial or non-financial competing interests. 

\section*{Author contribution}
CCo, LP and VB conceived the initial idea.
CCa and LC expanded the concept and developed the code.
LP, VB, CCa and LC developed the theoretical part. LC carried out the numerical analysis.
All the authors contributed to data analysis, figure preparation, 
and to manuscript writing.

\section*{Code availability}
The code used for this study is available in GitHub~\cite{Code}.
At the same link is available the code for designing the optimal network
for arbitrary spatial location of users.

\section*{Data availability}
The datasets used and/or analysed during the current study are
available from the corresponding author on reasonable request.

%\bibliography{qn-opti-biblio}{}

%\bibliographystyle{prsty_no_etal}

\end{document}